\documentclass{aastex6}
\newcommand{\ee}{$\epsilon$ Eri}
\begin{document}

\title{Radio Emission from the Exoplanetary System $\epsilon$ Eridani}

\author{T. S. Bastian\altaffilmark{1}, J. Villadsen\altaffilmark{1}, A. Maps\altaffilmark{2},  G. Hallinan\altaffilmark{3}, A. J. Beasley\altaffilmark{1}}
\affil{$^1$National Radio Astronomy Observatory, 520 Edgemont Road, Charlottesville, VA 22903, USA \\ 
$^2$Old Dominion University, Department of Physics, 4600 Eldhorn Avenue, Norfolk, VA 23529, USA\\
$^3$Caltech, Astronomy Department, 1200 E. California Boulevard, Pasadena, CA, 91125, USA}
\email{tbastian@nrao.edu}
\shorttitle{Radio Emission from \ee}
\shortauthors{Bastian et al.}

 
\begin{abstract}

As part of a wider search for radio emission from nearby systems known or suspected to contain extrasolar planets $\epsilon$ Eridani was observed by the Jansky Very Large Array (VLA) in the 2-4 GHz and 4-8  GHz frequency bands. In addition, as part of a separate survey of thermal emission from solar-like stars, \ee\ was observed in the 8-12 GHz and the 12-18 GHz bands  of the VLA. Quasi-steady continuum radio emission from \ee\ was detected in the three high-frequency bands at levels ranging from 67-83 $\mu$Jy. No significant variability is seen in the quasi-steady emission. The emission in the 2-4 GHz emission, however, is shown to be the result of a circularly polarized (up  to 50\%) radio pulse or flare of a few minutes duration that occurred at the beginning of the observation. We consider the astrometric position of the radio source in each frequency band relative to the expected position of the K2V star and the purported planet. The quasi-steady radio emission at frequencies $\ge \!8$ GHz is consistent with a stellar origin.  The quality of the 4-8 GHz astrometry provides no meaningful constraint on the origin of the emission. The location of the 2-4 GHz radio pulse is $>2.5\sigma$ from the star yet, based on the ephemeris of Benedict et al. (2006),  it is not consistent with the expected location of the planet either. If the radio pulse has a planetary origin, then either the planetary ephemeris is incorrect or the emission originates from another planet. 

 \end{abstract}

\keywords{stars: solar-type -- starts: activity -- planets and satellites: detection -- radio continuum: stars, planetary systems}

\section{Introduction} \label{sec:intro}

The detection and characterization of extrasolar planets, or exoplanets, has been a burgeoning field of study for more than two decades. At present, nearly $\approx\!3600$ confirmed planets have been detected, including nearly 600 systems containing multiple planets (see, for example, the NASA Exoplanetary Archive\footnote{http://exoplanetarchive.ipac.caltech.edu/}). Detections of exoplanets have relied on a variety of techniques: Doppler reflex or radial velocity measurements, transits, microlensing, and direct imaging. Absent from this list to date are direct detections of radio emission from exoplanets although there are good reasons to believe that magnetized exoplanets may be powerful radio emitters (Winglee et al. 1986; Bastian et al. 2000; Zarka et al. 2001). Jupiter is a useful prototype -- it emits intense coherent radio emission as a result of the cyclotron maser instability (CMI; see Treumann 2006 for a review) where fast electrons ($\sim 10$ keV) interact with Jupiter's polar magnetic fields to produce intense, circularly polarized radiation close to the local electron gyrofrequency $\nu_{Be}=eB/2\pi m_e c=2.8 B$ MHz, where $B$ is the magnetic field, $e$ is the electron charge, and $m_e$ is the electron mass (cgs units). Numerous searches for CMI emission from exoplanetary systems have been undertaken (Zarka et al. 2015) but most have been done at low frequencies  (a few $\times 100$ MHz or less) because Jupiter, with its polar magnetic fields of up to 14 G, produces CMI radio emission at frequencies $\le 40$ MHz. With the assumption that Jupiter-like exoplanets may have comparable, or at most a few times Jupiter's magnetic moment, one might expect by analogy radio emission ranging from a few times 10 MHz to perhaps a few times 100 MHz. The intriguing discovery of nonthermal radio emission from brown dwarfs at frequencies of several GHz (Berger et al. 2001), subsequently identified as auroral CMI emission (Hallinan et al. 2008), calls this assumption into question. Brown dwarfs are substellar objects that can be comparable to Jupiter in terms of size and rotation (see, e.g, Luhman 2012 for a review), but can have substantially larger magnetic fields than Jupiter -- as high as several kG (e.g., Hallinan et al. 2015). The distinction between brown dwarfs and giant planets is somewhat blurred as their mass domains overlap (Chabrier et al. 2014). Christensen et al. (2009) developed a scaling law based on the energy flux available for generating magnetic fields as the determinant of the magnetic field strength, applicable to both planetary dynamos and those of rapidly-rotating low-mass stars. Reiners \& Christensen (2010) subsequently applied  a modified version of the proposed scaling relation to a selection of exoplanets and brown dwarfs, showing that young exoplanets with masses $<13 M_{Jup}$ may have fields ranging from 100-1000 G. In view of these developments, a survey of nearby systems known or suspected of containing one or more exoplanets was initiated that uses the Jansky Very Large Array (VLA) to search the 1-8 GHz frequency range for CMI emission from exoplanets with a sensitivity of 10-20 $\mu$Jy. With CMI emission produced near $\nu_{Be}$ this frequency range corresponds to planetary magnetic field strengths of 360-2850 G, one to two orders of magnitude greater than that of Jupiter. There are approximately 50 systems within 20 pc containing nearly 100 exoplanets that are observable by the VLA. The 1-8 GHz frequency range is broken into 1-2 GHz, 2-4 GHz, and 4-8 GHz frequency bands. The survey is partially complete in the 1-2 GHz and 4-8 GHz bands and complete in the 2-4 GHz band. Results of the full survey will be published elsewhere. 

Among the objects surveyed was $\epsilon$ Eridani ($\equiv$ HD22049), a young (360-720 Myr; Janson et al. 2015), nearby solar-like star: a K2 dwarf with a mass of 0.82 M$_\odot$, a radius of 0.74 R$_\odot$, and an effective temperature of 5084 K. At a distance of only 3.2 pc, \ee\  has been a subject of scientific interest for many years, as well as a frequent subject of science fiction, television programming, and films. It was a target for Project Ozma (Drake 1961), an early radio search for extraterrestrial intelligence, as well as NASA's High Resolution Microwave Survey (Henry et al. 1995) and later, the privately funded Project Phoenix (Tarter 1996). It is known to be chromospherically active (Baliunas et al. 1981; Noyes et al 1984), to possess an active X-ray corona (Johnson 1981), and to display a magnetic activity cycle (Baliunas et al. 1995; Metcalfe et al. 2013; Saar \& Brandenburg 1999;  Jeffers et al 2017). Zeeman broadening observations suggest a mean, surface-averaged magnetic field of $165-190$ G (R\"uedi et al. 1997; Lehman et al. 2015). The coronal X-ray luminosity is $L_x\sim 10^{28.5}$ erg s$^{-1}$ (Johnson 1981; Schmitt et al. 1996) whereas that of the Sun ranges from $10^{26.8}-10^{27.9}$ ergs s$^{-1}$ over a solar cycle (e.g., Judge et al. 2003). The emission measure distribution is highly reminiscent of solar active regions, with the bulk of the coronal material at a temperature of $3\times 10^6$ K but with an X-ray filling factor of order unity (Drake et al 2000). Finally, \ee\ shows a mass loss rate that is perhaps 30 times that of the Sun (Wood et al. 2002). To summarize, the K2V star may be characterized as a ``young Sun" that displays levels of chromospheric and coronal activity that are more vigorous than those seen on the Sun. 

With the discovery of a substantial debris disk around \ee\ (Greaves et al. 1998), the detection of a Jupiter-sized planet (\ee\ b) orbiting the K2V star with a semi-major axis of 3.4 AU (Hatzes et al. 2000; Benedict et al. 2006), and the suggestion of a second planet at 40 AU (\ee\ c; Quillen \& Thorndike 2002), interest has broadened into consideration of \ee\ as a young solar system.  The cold debris disk has been studied extensively in recent years (e.g., Greaves et al. 2014; MacGregor et al. 2015, Chavez-Dagostino et al. 2016; Booth et al. 2017). It is located $\approx 64$ AU from the star and has a width of $\sim 20$ AU. Backman et al. (2009) suggested that, in addition to the cold, outer debris disk two narrow dust disks may also be present at $\sim 3$ and 20 AU, and that up to three planets may be required to maintain the the two inner disks and the inner edge of the outer broad disk.  More recently, Su et al. (2017) consider the origin of a warm inner disk and conclude that one (3-21 AU) or two (1.5-2 or 3-4 AU and 8-20 AU) dust-producing planetesimal belts are most likely responsible. Confirmation of the presence of one or more planets has been somewhat inconclusive. Combining more accurate radial velocity measurements with previous measurements Anglada-Escud\'e \& Butler (2012) were unable find solutions consistent with those of Benedict et al.  (2006). Zechmeister et al. (2013) were also unable to confirm the detection of \ee\ b. Anglada-Escud\'e \& Butler question the reality of \ee\ b, suggesting that the long term variability in radial velocity measurement may be instead the result of stellar activity. Howard \& Fulton (2016), however, found that extant Ca II H and K line observations of stellar variability do not correlate with radial velocity measurements.  In summary, the presence of \ee\ b and \ee\ c, their properties, and those of inner dust disks remain somewhat uncertain. We proceed on the assumption that a relatively young Jupiter-like planet may be present in the system a few AU from the K2V star. 

Zarka et al. (2015) have reviewed attempts to detect coherent radio emission from known or suspected exoplanets, including \ee, which has been a favored target for radio surveys in past years due to its proximity, the fact that the host star is solar-like, and more recently because it is suspected of harboring one or more planets that may interact with the substantial stellar wind. However, to date, no detections have been reported. Gary \& Linsky (1981) used the VLA, when it was far less sensitive than it is now, to observe \ee\ at 5 GHz. It was not detected at a $3\sigma$ limit of 0.32 mJy. George \& Stevens (2007) used the GMRT to observe \ee\ at 150 MHz but failed to detect it -- the $2.5\sigma$ limit is 7.8 mJy. Murphy et al. (2015) observed \ee\ with the MWA at 154 MHz as part of a survey of 17 southern exoplanetary systems. No detections were made and a $3\sigma$ upper limit of 7.7 mJy is reported. Here we report radio detections of \ee\ in four frequency bands spanning 2-18 GHz. In \S2 we describe the observations, data analysis, and results. In \S3 we discuss the results in the context of both planetary and stellar radio emission. We conclude in \S4.

\bigskip

\section{Observations and Results} \label{sec:obs}

The observations were acquired under three VLA programs, two associated with an ongoing survey of $\approx 50$ systems within 20 pc known to contain, or suspected of containing, one or more extrasolar planets; and one associated with a search for thermal radio emission from nearby solar-type stars. Selected results from the latter search have been reported by Villadsen et al. (2014). Observations of \ee\ performed as part of the exoplanets survey were made by the VLA in its C configuration. It was observed for 20 and 10 min in the 2-4 GHz and 4-8 GHz frequency bands, respectively. 3C48 was used as the flux and bandpass calibrator and J0339-0146 was the phase calibrator in both cases. The phase calibrator was observed only once in each case, just prior to the observation of \ee.  

Observations of the 8-12 GHz and 12-18 GHz frequency bands were made by the VLA in the D configuration, used 3C147 as the flux and bandpass calibrator; J0339-0146 was again the phase calibrator. The high frequency observations were of longer duration than the low-frequency observations with approximately 60 min and 90 min on source for the 8-12 GHz observations on two successive days. Approximately 60 min were spent on source for the 12-18 GHz observations. The phase calibrator was observed every 15 min for the 8-12 GHz and 12-18 GHz observations. All data sets were carefully edited to remove bad data and spurious signals due to radio frequency interference (RFI). The fractional bandwidth lost to RFI can be substantial in the compact VLA array configurations, especially at low frequencies: e.g., nearly 50\% of the 2-4 GHz spectral band was flagged as a result of RFI. A log of the observations and results is given in Table 1. Each data set was mapped out to the first null of the primary beam in order to identify and ``clean" the sidelobes of background sources. Self-calibration yielded only minor (few percent) improvement in the signal-to-noise ratio of sources within the field because the 2-4 GHz and 4-8 GHz data were low-frequency snapshots. Self-calibration of the higher frequency bands was marginal to impossible because there was insufficient flux in the field. For the purposes of astrometry (\S2.2) self-calibration is in any case to be avoided. 

\begin{deluxetable}{lcccccr}
\tablecaption{Observing Log and Results}
\tablehead{
\colhead{VLA} & \colhead{Observation} & \colhead{UT Time}  & \colhead{Band} & \colhead{$\theta_M/\theta_m/\phi$} & \colhead{S$_\nu$} & \colhead{$\rho_c$}\\
\colhead{Program} & \colhead{Date} & \colhead{Range} & \colhead{(GHz)} & \colhead{(asec/asec/deg)} & \colhead{($\mu$Jy)} & $(\%)$}
\startdata
16A-078 & 2016-Mar-01 & 00:51-01:11 & 2-4 &8.9/5.7/9.9& $<65^{a}$ & --- \\
&&&&& $940\pm93^b$ & 30-50\\
15B-326 & 2016-Jan-21 & 02:17-02:27 & 4-8  &20.1/3.7/73.5& 83$\pm$16.6 & $<50$\\
13A-471 & 2013-May-18 & 20:34-21:44 & 8-12 &9.2/4.4/55.2& 66.8$\pm$3.7 & $<14$\\
13A-471 & 2013-May-19 & 17:01-18:40 & 8-12 &7.2/5.2/-76.3& 70.3$\pm$2.7 & $<10$\\
13A-471 & 2013-Apr-20 & 22:24-23:27 & 12-18 &5.8/3.1/81.4& 81.2$\pm$6.6 & $<20$\\
\enddata
\tablenotetext{a}{ 2.5$\sigma$ upper limit for quasi-steady emission}
\tablenotetext{b}{ peak flux density of radio flare}
\end{deluxetable}

\begin{figure}[ht!]
\figurenum{1}
\begin{center}
\includegraphics[width=6in]{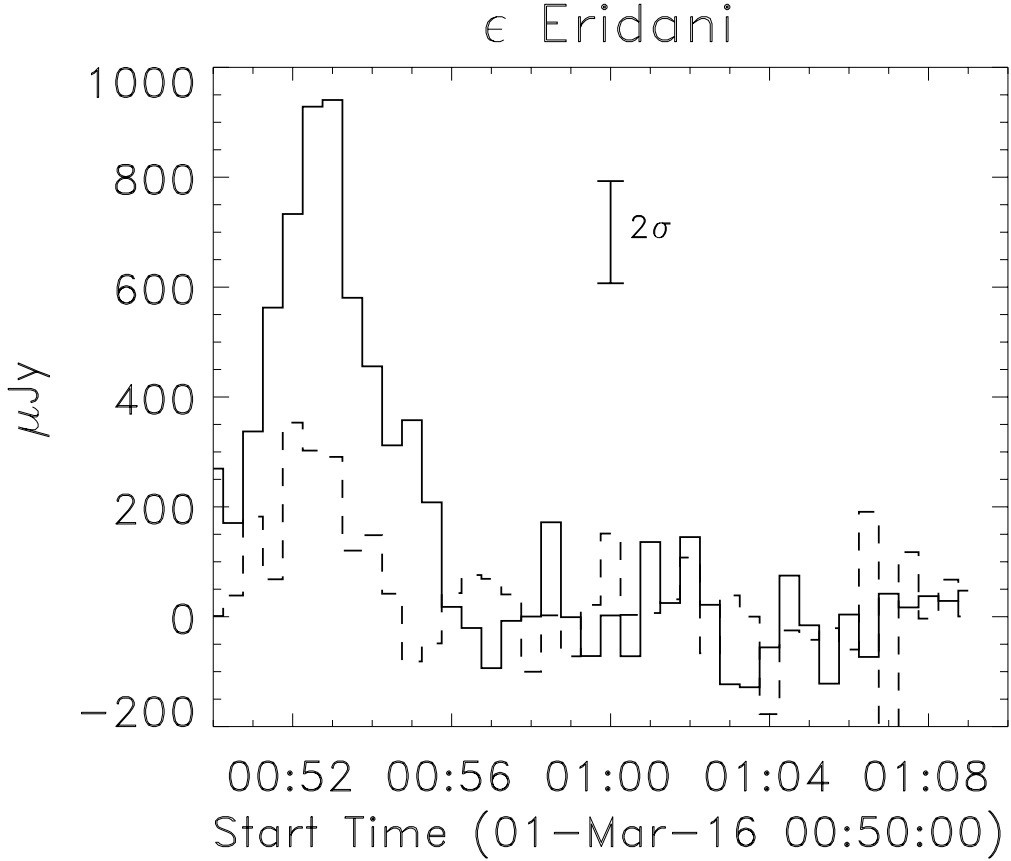}
\caption{The time variation of the the radio flux density (Stokes $I$ - solid line; Stokes V - dashed line) of a radio flare observed on \ee\ in the 2-4 GHz band. }
\end{center}
\end{figure}

\subsection{Continuum Radio Detections}

\ee\ was detected in total intensity (Stokes $I$) in all four frequency bands observed, as detailed in Table 1. No detection is made of steady continuum emission in the 2-4 GHz frequency band (see below), suggesting that the spectrum rolls off in or below the 4-8 GHz band. The most sensitive observations were in the 8-12 GHz band on 2013 May 18-19, sufficient to map the source at 6 frequencies across the band. We find that the spectrum is flat to within the noise. However, no specific conclusions can be drawn regarding the broadband spectrum of the steady continuum emission from band to band because they were individually observed months to years apart. 

Each band was assessed for time variability following the approach outlined by Kestevan et al. (1977) and employed by many other surveys (see, e.g., Swinbank et al. 2015, and references therein). Specifically,  for a time series of $n$ observations, the statistic $x_\nu=\sum_{i=1}^{n}{{(S_i-\bar{S})^2}/{\sigma_i^2}}$ is formed, where $\bar{S}$ is the weighted mean flux density. In the absence of source variability, $x_\nu$ is distributed as $\chi^2$ with $n-1$ degrees of freedom. The source flux density is taken to be variable if the probability $p(x_\nu)$ of exceeding $x_\nu$ by chance is $<0.1\%$, and is non-variable otherwise. In order to perform the analysis it was necessary to remove the influence of time variable sidelobes from background sources. To do so, the clean components associated with all background sources in a give frequency band were subtracted from the database leaving only radio emission from \ee. The data were then binned in 60s intervals and the analysis described was performed. We characterize the 4-8, 8-12, and 12-18 GHz bands  as being quasi-steady to the extent that they show no evidence of significant variability during the course of an observation. Circularly polarized emission (Stokes $V$) was not detected in the quasi-steady emission from \ee\ although the significance is low in the 4-8 GHz band. If the limit to Stokes $V$ is taken to be 2.5$\sigma$ in each band, the upper limits to the degree of circular polarization $\rho_c=V/I$ are as shown in Table 1.

The 2-4 GHz band is another matter. An impulsive radio event  - a pulse or a flare - was serendipitously observed during the first few minutes of the observation of \ee. After detecting the source at a level of  $203\ \mu$Jy over the 20 min duration of the observation, the source was mapped every 30~s for the duration of the observation in both Stokes I and V. The source flux density was already elevated when the observation began and rose to a maximum of nearly $S_\nu \approx 1$ mJy over the course of 3 min, after which it declined and was no longer detectable after another 3 min. Integrating over the post-flare period, \ee\ was not detected although the sensitivity of the post-flare period was relatively poor, yielding a 2.5$\sigma$ upper limit of 65 $\mu$Jy.  The flare displayed significant circularly polarized emission, reaching a maximum somewhat earlier than the maximum in total intensity. At the time of maximum Stokes $V$ the degree of circular polarization was $\approx 50\%$, declining to $30\%$ at the time of peak total intensity. The spectrum of the emission across the 2-4 GHz band is not well constrained as a result of significant gaps in the frequency coverage resulting from RFI flagging. However, averaging over frequencies between 2.5-3 GHz and 3-3.5 GHz, and integrating the data for 1 min at the time of the flare maximum, yielded a spectral index of $\sim\!1.5$; i.e., a spectrum that increases with frequency.

\subsection{Source Locations}

If the orbital parameters of Benedict et al. (2006) are correct, the angular separation between \ee\ b and the K2V primary is 1.8" at apastron; for the two epochs presented here, roughly 2013.38 and 2016.11, the expected angular separation would be 0.93" and 1.57", respectively. To assess source positions and their uncertainties in each frequency band we  follow Condon (1997) and Condon et al. (1998). The uncertainties parallel to the FWHM major axis ($\theta_M$) and minor axis ($\theta_m$) of the synthesized beam depend on the effective signal-to-noise ratio (SNR) as

\begin{equation}
{{\sigma_M^2}\over{\theta_M^2}}={{\sigma_m^2}\over{\theta_m^2}}\approx {1\over{8\ln 2}} (SNR)^{-2} \nonumber
\end{equation}

\noindent The uncertainties in right ascension ($\sigma_\alpha$) and declination ($\sigma_\delta$) are then

\begin{eqnarray}
\sigma_\alpha^2 \approx \epsilon_\alpha^2+\sigma_M^2 \sin^2\phi +\sigma_m^2 \cos^2\phi \nonumber \\
\sigma_\delta^2 \approx \epsilon_\delta^2+\sigma_M^2 \cos^2\phi +\sigma_m^2 \sin^2\phi \nonumber
\end{eqnarray}

\begin{figure}[ht!]
\figurenum{2}
\begin{center}
\includegraphics[angle=90,width=7in]{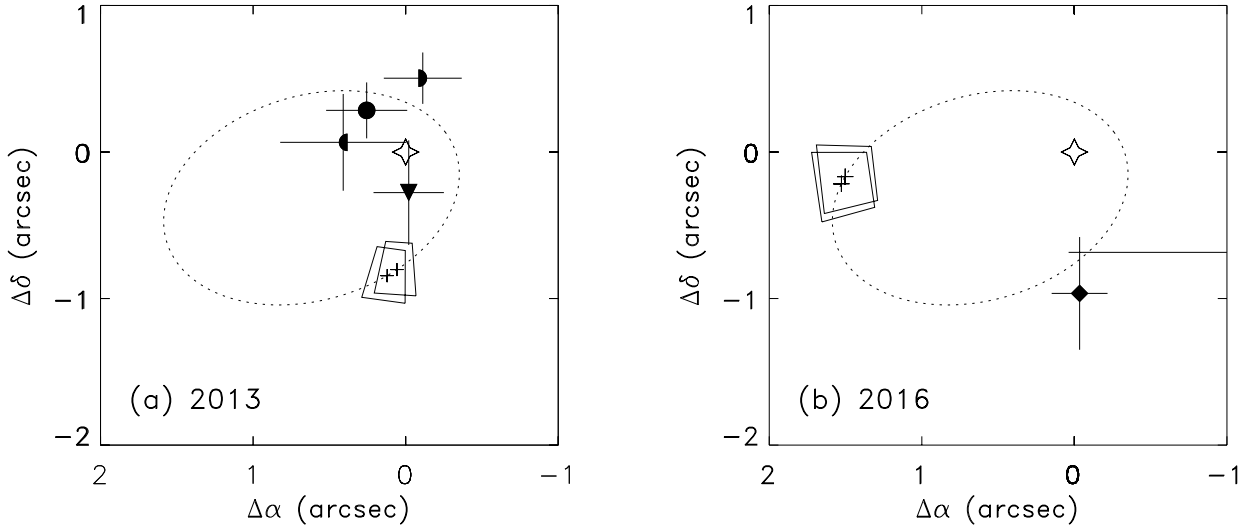}
\caption{a) The apparent locations of the radio sources detected toward \ee\ relative to the location of the primary K2V star and the purported planet. The open star symbol represents the location of the K2V star. The filled semi-circles represent the location of the 8-12 GHz source on 2013 May 18 (left-fill) and May 19 (right-fill), and the filled circle represents the position determined from the combined data; the inverted triangle represents the location of the 12-18 GHz source. The orbit of \ee\ b according to Benedict et al. (2006) is shown as a dotted line and the location of the planet is shown as plus signs. The trapezoidal boxes represent uncertainties in the location of the planet. b) The same for the observations made in 2016. The diamond represents the 2-4 GHz source; the apparent position of the 4-8 GHz source is outside the box plotted, with just its error bar in $\Delta\alpha$ visible. The location of the \ee\ b and the associated uncertainties are again shown using the ephemeris of Benedict et al.  }
\end{center}
\end{figure}

\noindent where $\epsilon_\alpha$ and $\epsilon_\delta$ are calibration errors taken to be 0.1" and the synthesized beam parameters $\theta_M$, $\theta_m$, and $\phi$ are given in Table 1. Figure 2 summarizes the apparent positions and uncertainties of the radio detections relative to the position of the K2V star. Also shown is the position of the purported planet \ee\ b using the ephemeris of Benedict et al. (2006) at the epochs during which the radio observations were made.  Uncertainties in the location of the planet follow from Janson et al. (2007). The 8-12 GHz, and 12-18 GHz sources, observed in 2013, are consistent with radio emission from the K2V primary. The apparent location of the 4-8 GHz source is poorly determined as a result of the low SNR of the observation and the low elevation of the source at the time it was observed in 2016: $\Delta\alpha=-1.82"\pm 1.78,\ Delta\delta=-0.68\pm0.54"$), placing it outside the domain plotted in Fig. 2 but, formally, within 1$\sigma$ of the K2V star.  The location of the 2-4 GHz pulse emission, however, is more than 2.5$\sigma$ from the star but is even farther from the nominal location of \ee\ b . If the planetary ephemeris of Benedict et al. is correct, the radio pulse is clearly inconsistent with a planetary origin. We conclude that either i) there is a source of systematic error that is unaccounted for in the 2-4 GHz astrometry and the source is associated with the K2V star, or ii) the planetary ephemeris of Benedict et al. is not correct and the emission is associated with a planet, or iii) the 2-4 GHz emission is associated with a planet different from that identified by Benedict et al. 

\section{Discussion} \label{sec:disc}

Radio emission from \ee\ has been detected over a wide range of dates and frequency bands. For the purposes of discussion we distinguish between the quasi-steady emission observed in the high-frequency bands and the impulsive, circularly polarized event detected in the 2-4 GHz band. We have concluded that the quasi-steady radio emission at frequencies $\ge 8$ GHz originates on the K2V star; for the purposes of discussion, we assume that the 4-8 GHz quasi-steady emission does, too. We consider both a stellar origin and a planetary origin for the radio pulse observed in the 2-4 GHz frequency band. 

\subsection{A Stellar Origin}

\subsubsection{Quasi-steady Continuum}

It is well established that for magnetically active stars the radio luminosity $L_R$ near 5 GHz is correlated with the soft X-ray luminosity $L_X$ through the so-called G\"udel-Benz relation: $L_x/L_R\approx 10^{15.5}$ Hz (Benz \& G\"udel 1994; G\"udel et al. 1995). The soft X-ray emission is attributed to thermal coronal plasma and the radio emission is attributed to nonthermal gyrosynchrotron emission. While the correlation applies to solar flares, it does not apply to relatively magnetically quiet stars like the Sun (for which $L_X/L_R\sim 10^{17}$). The radio luminosity of \ee\ in the 4-8 GHz band is $L_R\sim 10^{12}$ erg s$^{-1}$ Hz$^{-1}$ so that $L_X/L_R \sim 10^{16.5}$ Hz, suggesting that $L_R$ is under-luminous relative to active stars, and consistent with the idea that while \ee\ is younger and more active than the Sun it is not an ``active star". We argue here that nonthermal radio emission is not required to account for the observed quasi-steady emission.

Expressing the flux density in terms of brightness temperature $T_B$ we have $T_B=3.8\times 10^5 S_{\mu{Jy}} \nu_9^{-2} (r_S/R_\ast)^{-2}$ where $S_{\mu Jy}$ is the flux density in $\mu$Jy at frequency $\nu_9$ GHz, and $r_S$ is the source radius in units of the K2V stellar radius $R_\ast$. Taking $r_S\approx R_\ast$ results in $T_B <2.7\times 10^6$ K (2-4 GHz); $T_B=8.8 \times 10^5$ K (4-8 GHz); $T_B=2.6\times 10^5$ K (8-12 GHz); and $T_B=1.4\times 10^5$ K (12-18 GHz). The brightness temperatures can be understood if we assume they are the result of optically thick contributions from both the stellar chromosphere and the low corona.  Assuming that the chromospheric temperature is comparable to that of the Sun (Jordan et al. 1987) we have $T_{ch} \approx 10^4$ K. If the radio emission from \ee\ was the result of chromospheric emission alone, with $T_B \approx T_{ch}$, it would be negligible, yielding only $S_\nu$ =  0.24, 0.96, 2.65, and 5.97 $\mu$Jy in the 2-4, 4-8, 8-12, and 12-18 GHz bands, respectively. Taking the coronal brightness temperature to be $T_c\sim 3\times 10^6$ K, the temperature at which the differential emission measure of of X-ray emitting material peaks (Drake et al. 2000), and assuming the observed brightness temperature is $T_B = f_c T_c + (1-f_c) T_{ch}$, where $f_c$ is the filling factor of material with a brightness temperature of $T_c$ on the stellar disk, we have $f_c<90\%$ in the 2-4 GHz band and $f_c=29\%$ (4-8 GHz), $f_c=7-8\%$ (8-12 GHz), and $f_c=4\%$ (12-18 GHz). 

Two sources of opacity are likely relevant to radio emission in the corona of \ee: thermal free-free absorption and thermal gyroresonance absorption. For the former the absorption coefficient $\kappa_{ff} \propto \nu^{-2}T^{-3/2} n_e^2$, whereas for the latter, $\kappa_{gr} \propto n_e T_e^\alpha B^\beta$ with $\alpha, \beta >1$ (Dulk 1985). Hence, free-free absorption is favored for cooler, denser plasmas and low frequencies whereas gyroresonance absorption favors hotter plasmas and strong magnetic fields. Both play a role on the Sun above solar active regions (e.g., Gary \& Hurford 1994) with free-free absorption playing a dominant role for frequencies below 2-3 GHz.  At higher frequencies, where coronal free-free absorption is no longer relevant, magnetic fields may be sufficiently strong to render the solar corona optically thick to gyroresonance absorption at low harmonics of $\nu_{Be}$; that is, $\nu=s\nu_{Be}$. In the case of the Sun $s=2$ for the ordinary mode and $s=3$ for the extraordinary mode for typical coronal conditions (White \& Kundu 1997). Gyroresonance emission can be relevant on the Sun from $\sim 2-18$ GHz, the upper frequency limit determined by the maximum coronal magnetic field strength above sunspots. Brosius \& White 2006, for example, deduced a magnetic field strength of 1750 G in the low corona above a sunspot from its emission at 15 GHz. The filling factor of magnetic fields with a strength sufficient to render the corona optically thick to gyroresonance absorption at a given frequency varies on the Sun inversely with frequency: the fractional area above an active region at coronal brightness temperatures at high frequencies is significantly less than it is for low frequencies (see, e.g., Lee et al. 1998). We assume coronal conditions on \ee\ are similar to those above a solar active region (Drake et al. 2000) and that a significant fraction of the low corona of \ee\ is therefore optically thick to free-free absorption in the 2-4 GHz frequency band. For this to be the case, with a coronal temperature $T_c\sim 3\times 10^6$ K and a frequency of 3 GHz, $\tau_{ff}\gtrapprox 1$ for a column depth $n_e^2L \gtrapprox 2\times 10^{29}$ cm$^{-5}$. Taking $L$ to be comparable to the gravitational scale height, a mean coronal electron number density $n_e \gtrapprox 4\times 10^9$ cm$^{-3}$ is required. A higher column depth of coronal plasma would result in free-free absorption being relevant at yet higher frequencies. Assuming, however, that at frequencies greater than 2-4 GHz the corona of \ee\ is likely optically thin to free-free absorption, gyroresonance absorption predominates, suggesting that magnetic field strengths of 475 G to as high as 2700 G are present in the low corona with filling factors ranging from 29\% to 4\% producing gyroresonance emission in the  4-8 GHz and 12-18 GHz, respectively. These do not appear to be inconsistent with mean, disk-averaged magnetic field strength of $186\pm 47$ G measured at the photospheric level reported by Lehman et al. (2015). At even higher frequencies, the coronal is optically thin to both free-free and gyroresonance absorption. Using observations at 43 and 230 GHz MacGregor et al. (2015) find a disk-averaged brightness temperature of 13,000 K, consistent with free-free emission from the upper chromosphere. 

\subsubsection{A Stellar Flare}

The quasi-steady radio emission is consistent with optically thick radio emission at coronal brightness temperatures above active regions. As active region magnetic fields evolve they almost certainly result in magnetic energy release in the form of flares. Hence, the flare may have originated on the star, too. If we suppose that the source scale of the radio flare in the 2-4 GHz band was comparable to that of a large solar flare, with $r_S\sim 10^{10}$ cm, the peak flux density $S_{\mu Jy} \approx 940$ yields a brightness temperature $T_B\approx 10^9$ K which suggests that a nonthermal mechanism is responsible for the emission; specifically, nonthermal gyrosynchroton emission from mildly relativistic electrons interacting with the coronal magnetic field of the K2V star. For the purposes of illustration, we assume the bulk of the emission is marginally optically thick and that it is due to a power-law distribution of electrons $n(E)=A E^{-\delta}$ ($A$ a normalization constant). If the coronal magnetic field in the flare source is oriented $\sim 60^\circ$ to the line of sight, we find that for $\delta=4,5,6$ the bulk of the emission originates at a peak harmonic $s_{pk} =12, 20, 31$ (Dulk 1985) and so $B = 92, 54, 35$ G, respectively. If we instead assume that the coronal magnetic field is oriented $30^\circ$ to the line of sight we have $s_{pk} =8, 13, 21$ and $B = 136, 80, 51$. Qualitatively, a harder electron distribution and/or a line of sight more closely aligned with the magnetic field implies the need for somewhat stronger coronal magnetic fields. In addition, lower harmonics yield a higher degree of circular polarization for a marginally optically thick source. 

Alternatively, if the flare was the result of coherent CMI emission, a magnetic field in the source of $\nu=\nu_{Be}=700-1400$ G is implied (corresponding to the bandwidth of 2-4 GHz). The source size is likely such that $r_S\ll 10^{10}$ cm, leading to a brightness temperature $T_B\gg 10^9$ K. The presence of such magnetic fields is already implied by the quasi-steady radio emission and so conditions on the star may be favorable for the production of CMI emission, as they are on the Sun. A problem for CMI emission in coronal environments is the escape of the radiation. Since it is produced at $\nu=\nu_{Be}$ it is subject to strong gyroresonance absorption in harmonic layers that the radio radiation must necessarily traverse to escape the corona (Melrose \& Dulk 1981) although there may be special circumstances under which escape of the radiation is possible (Zaitsev et al. 2005). 

It is interesting to note that had the flare been observed from 1 AU it would have have had a maximum of 44000 Solar Flux Units\footnote{1 SFU = $10^4$ Jy}. Such a flare on the Sun at this frequency resulting from incoherent gyrosynchrotron would be rare, but not unprecedented (e.g., Nita et al. 2002). Large solar flares are associated with fast coronal mass ejections, coronal and interplanetary shocks, and the acceleration of energetic particles. Such a powerful flare could produce secondary emissions resulting from shocks (e.g., type-II-like radio bursts - see Crosley et al. 2016) and/or interaction with a planetary magnetosphere. If a magnetized, Jupiter-sized planet is in fact present a few AU from the K2V star, it is exposed to a strong stellar wind ($\sim 30\times$ solar; Wood et al. 2002), and is also likely exposed to fast coronal mass ejections and associated shocks, and the enhanced electromagnetic and hard particle radiation that flares and CMEs produce. 

\subsection{A Planetary Origin for the Radio Pulse}

The apparent location of the polarized radio pulse observed in the 2-4 GHz band is not well associated with either the K2V star or the planet, assuming the planetary ephemeris is correct.  Either the astrometry is incorrect, or the planetary ephemeris is incorrect, or another planet is present in the system. Are the properties of the radio pulse consistent with a planetary origin? It is well-established that recurrent CMI emission occurs at GHz frequencies on brown dwarfs and that it is most likely auroral in nature (e.g., Hallinan et al. 2008). The radio flare from \ee\  is only minutes in duration and has a degree of circular polarization of up to 50\%. By way of comparison, recurrent auroral radio emission attributed to CMI emission observed from a sample of five L and T dwarfs by Kao et al. (2016) shows that the ``pulsed" component displays degrees of circular polarization ranging from 28\% to as high as 97\%;  pulses may even be unpolarized (Hallinan et al. 2008; Lynch et al. 2015). The pulse durations are of order minutes. The radio flare seen from \ee\ is not inconsistent with these properties. As noted when considering a stellar origin, CMI emission implies that the magnetic field strength in the source ranges from 700-1400 G for emission spanning the 2-4 GHz frequency band and the planet would therefore possess a (polar) magnetic field of this magnitude. Unlike the case of a stellar CMI source, emission from a planet would not necessarily have difficulty escaping the source if the density of thermal plasma in the planetary magnetosphere is very low or if the thermal plasma is confined to a limited volume - an equatorial torus, for example, which is the case for Jupiter (e.g., Badman et al. 2015). Models of CMI emission from strong (several kG) dipolar magnetic fields have been explored elsewhere (Yu et al. 2011; Kuznetsov \& Vlasov 2012; Kuznetsov et al. 2012) and will not be pursued here. We note that Reiners \& Christensen (2010) estimate that a polar magnetic field of just 19~G is expected on \ee\ b based on their scaling law, far below the requirement for a planetary CMI source in \ee. However, at least in the case of  brown dwarfs, there may be significant deviations from the proposed scaling (Kao et al. 2016).

Longer term monitoring is necessary to determine whether the flare-like emission from \ee\ is recurrent or not. If it does recur on a period that is significantly different from the 11.2~d rotational period of the K2V star, the case for a planetary origin would be strong. The planet would be unusual to the extent that it would be the first to show evidence for a very large magnetic field, similar in magnitude to those inferred for brown dwarfs. If the radio source is indeed associated with the purported planet \ee\ b,  the radio emission may be ultimately powered by interaction between the planetary magnetosphere and the stellar wind (Zarka et al. 2001). Coronal mass ejections from the star may also play a role in powering or modulating planetary CMI radio emission, as could the presence of one or more satellites around the planet.

\bigskip\bigskip

\section{Summary and Conclusions} \label{sec:conc}

Radio emission has been detected from \ee\ in four radio frequency bands spanning 2-18 GHz.  Quasi-steady radio emission was detected at levels ranging from approximately 67-83 $\mu$Jy in the three high-frequency bands. The quasi-steady emission was unpolarized within sensitivity limits and showed no significant time variability during the course of a given observation. The quasi-steady radio emission from \ee\ is consistent with optically thick emission at coronal temperatures from the K2V star as a result of free-free absorption at lower frequencies where the filling factor of dense, active region plasma is high. It is attributed to thermal gyroresonance emission at frequencies $>3-4$ GHz in regions where the coronal magnetic field is sufficient to render the coronal optically thick. The filling factor of magnetic fields at low coronal heights varies from 29\% to 4\% for magnetic fields ranging from 475 G to 2700 G, respectively. 

A radio pulse or flare was detected from $\epsilon$ Eri in the 2-4 GHz frequency band. The origin of the flare may be the K2V star which is known to be magnetically variable and to have magnetic fields consistent with those inferred from the radio observations of the quasi-steady emission. We have considered incoherent gyrosynchrotron emission and coherent CMI emission as possible mechanisms for the stellar flare although the CMI mechanism may require special conditions for the emission to avoid strong absorption by the overlying corona. We have also considered whether the observed emission could plausibly originate from the purported exoplanet, \ee\ b, a Jupiter-sized planet orbiting the K2V star. We found that the radio source is not consistent with a planetary origin if the ephemeris of Benedict et al. (2006) is correct. If the emission is from \ee\ b, either the ephemeris of Benedict et al. is incorrect or the source is another planet in the system. If the planet has a kG magnetic field it could be responsible for the polarized pulse via CMI emission. If such a planet is present, it would likely interact strongly with the stellar wind from the star and with ``space weather" events driven by flares and coronal mass ejections from the star. Radio monitoring of \ee\ is required to definitively establish the origin of radio variability in the 2-4 GHz frequency band. 

\acknowledgments
The National Radio Astronomy Observatory is a facility of the National Science Foundation operated under cooperative agreement by Associated Universities, Inc. A. Maps was supported by NSF grant AST-1358169 to Associated Universities, Inc., in support of the Research Experience for Undergraduates summer student program. This work utilized data acquired by VLA programs 13A-471, 15B-326, and 16A-078. We thank the referee for constructive comments that improved the paper.



\end{document}